\documentclass[aps,prl,twocolumn,superscriptaddress,floatfix]{revtex4}
\usepackage{graphicx}

\begin{document}


\title{Finite-Size Scaling and Universality of the Thermal Resistivity of Liquid $^4$He near $T_\lambda$}


\author{Daniel Murphy}
\author{Edgar Genio}
\author{Guenter Ahlers}
\affiliation{Department of Physics and iQUEST,\\ University of
California, Santa Barbara, CA  93106}
\author{Fengchuan \surname{Liu}}
\author{Yuanming \surname{Liu}}
\affiliation{Jet Propulsion Laboratory, California Institute of Technology, Pasadena, CA 91109}

\date{\today}

\begin{abstract}
We present measurements of the thermal resistivity $\rho(t,P,L)$ near the superfluid transition of $^4$He at saturated vapor pressure and confined in cylindrical geometries with radii $L=0.5$ and 1.0 $\mu$m ($t\equiv T/T_\lambda(P) - 1$). For $L = 1.0 \mu$m measurements at six pressures $P$ are presented.  At and above $T_\lambda$ the data are consistent with a universal scaling function $F=\left(L/\xi_0\right)^{x/\nu}\left( R/R_0\right)$,  $X= \left(L/\xi_0\right)^{1/\nu}t$ valid for all $P$ ($R_0$ and $x$ are the pressure-dependent amplitude and effective exponent of the bulk resistivity $R$ and $\xi = \xi_0 t^{-\nu}$ is the correlation length).

\end{abstract}


\maketitle

The modern theory of critical phenomena \cite{fisher:1998:1} predicts that continuous phase transitions belong to distinct universality classes which are determined by such general properties of the system as the number of degrees of freedom of the order parameter and the spatial dimensionality. Within a given class, exponents and amplitude ratios are identical (i.e. universal) for all members and independent of irrelevant variables. An example of an irrelevant variable is the pressure $P$ of a liquid helium sample at which measurements near the superfluid transition temperature $T_\lambda$ are made. Within a given universality class, the dependence of many properties upon certain parameters can be represented by scaling functions which are the same for all systems. The present paper is an experimental study of the scaling function which describes the effect of confinement in a cylindrical geometry with radius $L$ on critical properties. For static properties this finite-size scaling has been studied by a number of precise experiments. For example, the heat capacity near the superfluid transition of $^4$He has been measured for confinement sizes which vary by a factor of 500, and the data to a large extent can be collapsed upon a unique function when properly reduced \cite{mehta:g:1997:1,lipa:etal:2000:1}. Even for static properties, however, we are not aware of any measurements which test the universality of finite-size scaling.  For transport properties there are, to our knowledge, no prior experiments which test scaling and universality for finite-size effects. There has been only one experiment on the effects of confinement on a transport property \cite{kahn:a:1995:1,ahlers:1999:1}, namely the measurements of the thermal conductivity $\lambda$ of helium near $T_\lambda$ in cylindrical tubes. These measurements can be used to derive a scaling function which would be expected to be universal, but since they were performed only for the one value  $L = 1 \mu$m and only at saturated vapor pressure (SVP), they provided a test of neither finite-size scaling nor of the universality of the derived function. 

We present experimental results for the thermal resistivity $\rho(t,P,L) \equiv 1/\lambda(t,P,L)$ near $T_\lambda(P)$ of liquid $^4$He confined in cylinders of two different radii and at various pressures as a function of the reduced temperature $t \equiv T/T_\lambda - 1$.  The use of two confinement sizes allows us to directly test finite-size scaling, while the use of different pressures for one size provides a test of universality. For bulk helium $\rho(t,P,\infty)$ depends strongly on pressure \cite{tam:a:1985:1}, so that a comparison of an appropriate scaling function for $\rho(t,P,L)$ at different pressures provides a sensitive test of universality. These two aspects were tested in separate experiments: measurements as a function of $L$ were taken at SVP, and measurements as a function of $P$ were taken at a single confinement size $L=1.0\mu$m.

Theoretical predictions for $\lambda$ are still quite limited. Monte Carlo calculations give the shape of a scaling function, but only to  within a multiplicative factor \cite{nho:m:2001:1}. Within its precision this shape agrees well with the measurements of Ref.~\cite{kahn:a:1995:1}. Very recently, a one-loop renormalization group (RG)  calculation of $\lambda(t,P,L)$ for $t \ge 0$ and at SVP was carried out by T\"opler and Dohm \cite{topler:d:2002:1}, but at present there are no such calculations for $t < 0$ and for higher pressures.
Thus, in order to provide a broader framework for the analysis of our data, we use a phenomenological approach. We assume that the temperature and size dependence of $\rho$ are separable and that the size dependence is a function only of $L/\xi$ where $\xi = \xi_0t^{-v}$ is the correlation length: $\rho(t,P,L) = \rho(t,P,\infty)\tilde{F}\left({L}/{\xi}\right)$.  Since $\rho(t,P,\infty)$ goes to zero as $t$ does while $\rho(t,P,L)$ remains finite,  $\tilde{F}$ diverges at $t=0$.  To avoid this difficulty, we re-define the scaling function as $F(X) = \left({L}/{\xi}\right)^{x/\nu} \tilde{F}$ which avoids the divergence at $t = 0$. Consistent with experiment \cite{tam:a:1985:1},  we have written $\rho$ for bulk helium as a power law $\rho(t,P,\infty) = \rho_0 t^x$ with effective exponents $x(P)$  and amplitudes $\rho_0(P)$. We now have
\begin{equation}
F(X) = \left[\frac{L}{\xi_0(P)}\right]^{x/\nu}\left[\rho(t,P,L)/\rho_0(P)\right]
\label{eq:F}
\end{equation}
with
\begin{eqnarray}
X =& \left(\frac{L}{\xi_0}\right)^{1/\nu}t\nonumber \\
=& t/t_0,
\label{eq:X}
\end{eqnarray}
Note that $t_0$ is the temperature at which the correlation length grows to the size of the container, i.e. $\xi \simeq L$. The correlation length has a pressure-dependent amplitude  $\xi_0(P)$ and a universal exponent $\nu$. The values of $\xi_0, \rho_0,$ and $x$ are known from bulk measurements \cite{tam:a:1985:1} and are summarized in Table 1.

We used two different thermal conductivity cells.  One (Cell I) was described in detail elsewhere \cite{kahn:a:1995:1}. It was used for measurements of the resistivity as function of $P$ at $L=1.0\mu$m. It consisted of two cylindrical metal plates made of OFHC (Oxygen-Free High-Conductivity) copper separated by a stainless steel sidewall.  A glass microchannel plate (MCP) was epoxied to the inside of the sidewall, so that when assembled the liquid helium between the plates would be confined to the channels with little extraneous liquid between the endplates and the glass.  A small bulk thermal conductivity cell, called a ``lambda device'',  was attached to the bottom (hot) copper plate for the determination of $T_\lambda$ in bulk helium.  The bottom of the lambda device was $1.25$~cm below that of the confinement cell, and as a result the value of $T_\lambda$ had to be corrected for the hydrostatic pressure difference between the bottom of the lambda device and the middle of the MCP \cite{ahlers:1968:1,ahlers:1991:1}. Since the lambda device was attached to the bottom plate, it could only be used before and after a data acquisition sequence because the heat applied to it necessarily flowed through the confining cell.  The cell was filled through an overflow volume located on the top (cold) copper plate.

Cell~II, used for measurements of the resistivity as a function of $L$ at SVP, was designed for use with microchannel plates which were surrounded by a solid glass ring. Whereas the MCP in Cell I was epoxied into a stainless steel sidewall which in turn was sealed to the copper endplates with indium gaskets, the glass ring in the second type was directly sealed to the copper using indium.  A stainless steel sidewall was used as a spacer, but its length was chosen so that the different thermal expansion coefficients of the copper endplates and the stainless steel compensated each other. As a result the force applied to the microchannel plate was constant.  The cryogenic apparatus used with this cell design could accommodate three thermal conductivity cells, all of which were suspended from a common temperature-regulated platform.  One of these cells was a bulk conductivity cell constructed with an open glass ring. It served to locate $T_\lambda$ of the bulk fluid.  The vertical centers of the cells were nearly the same, so that the gravity correction mentioned for the first cell was greatly reduced.  The three cells were independent of each other, so that the bulk thermal conductivity could be measured at the same time as the thermal conductivity for two different confinement sizes.  The fill line entered the bottom of the cell, and the portion of the fill line located in the bottom plate was packed with $0.05~\mu$m alumina powder to suppress the superfluid transition. Thus the liquid helium contained in the bottom plate was always normal.  The fill line was connected to an overflow volume on the shield stage, which was maintained a few mK above $T_\lambda$.

Saturated vapor pressure was maintained in both sets of experiments by partially filling the overflow volumes.  Pressures other than SVP were reached in Cell~I using a ``hot volume'' \cite{mueller:a:p:1976:1}, a separate thermal stage filled with fluid whose temperature was controlled to regulate the pressure in the thermal conductivity cell.  The pressure was measured using a capacitative strain gage \cite{straty:a:1969:1} mounted on the top of the cell.  The fluid in the cell and the hot volume was isolated from the rest of the cryostat by a normally-closed low-temperature valve.

The resistivity $\rho$ was computed from the temperature difference $\Delta T$ which was measured across the fluid layer when a power $Q$ was applied to the bottom plate of a  cell: $\rho = ({A}/{d})\left[\left({Q}/{\Delta T} - C_W\right)^{-1} - R_b\right]$, where $A$ is the cross-sectional area of the fluid, $d$ is the spacing between the plates, $C_W$ is the parallel conductance of the stainless steel sidewall and the glass of the MCP, and $R_b$ is the boundary resistance between the copper endplates and the fluid.  The boundary resistance $R_b$ was measured far below $T_\lambda$, where the resistance of the fluid layer can be neglected.  The size of the correction is relatively small, and its temperature dependence near $T_\lambda$ \cite{kuehn:etal:2002:1} was neglected. The parameters $A/d$ and $C_W$  were obtained by fitting the measured  $\lambda(t,P,L)$ to the known  $\lambda(t,P,\infty)$ several mK above $T_\lambda$ where the effects of  confinement are negligible.  For Cell~I, the pressures at which measurements were made were chosen to match those for which prior measurements for bulk helium were available \cite{tam:a:1985:1}; the value $d/A=0.386$ so obtained was found to be independent of pressure and agreed with the value $0.39$ previously determined for this cell \cite{kahn:a:1995:1}.  The $0.5$ and $1~\mu$m data, taken with Cell~II, yielded $d/A=0.0781$ and $0.0605$ respectively.  All values for $d/A$ are in  good agreement with  values from gas flow-impedance measurements on and electron micrographs of the microchannel plates.  The values for $C_W$ for each size and pressure are shown in Table~\ref{tab:F0vsP} (for $P=11.25$~bar, there were no bulk conductivity data, and $C_W$ was obtained by interpolation between other pressures).  Each conductivity data point was assigned to the mean temperature $\bar T=T_{\textrm{top}}+\Delta T/2$, and a corresponding curvature correction \cite{tam:a:1985:1} was applied to correct for the use of a finite $Q$ and $\Delta T$.

\begin{figure}
 \includegraphics[width=8.5cm]{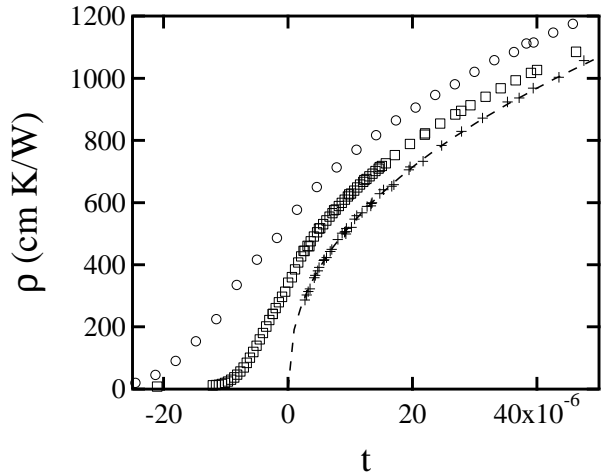}
 \caption{\label{fig:rhoL} Thermal resistivity versus reduced temperature at SVP for $L = 0.5 \mu$m (open circles) and 1.0$\mu$m (open squares). The plusses are bulk measurements (Ref.~\protect \cite{tam:a:1985:1}) and the dashed curve is a powerlaw fit to the bulk data.}
 \end{figure}

\begin{figure}
 \includegraphics[width=8.5cm]{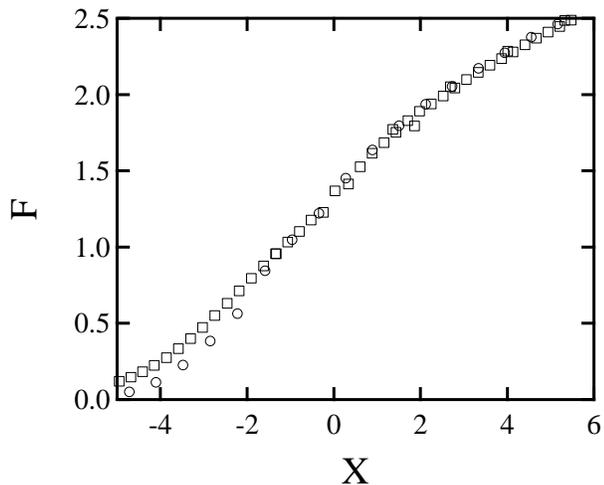}
 \caption{\label{fig:FvsXL} Scaling function $F$ versus scaling variable $X$ at SVP for $L = 0.5 \mu$m (open circles) and 1.0$\mu$m (open squares).}
 \end{figure}

The resistivity at SVP is plotted versus $t$ in Fig.~\ref{fig:rhoL} for two different values of $L$.  The data show the effect of confinement, with the smallest size showing the greatest rounding of the transition and the greatest increase of $\rho(t=0)$.  The scaling variable $F$ (Eq.~\ref{eq:F}) is plotted versus $X$ (Eq.~\ref{eq:X}) for the two sizes in Fig.~\ref{fig:FvsXL}.  Except perhaps for $X \alt -2$, the data collapse onto a single curve, thus supporting the concept of finite-size scaling. It is difficult to tell whether the small difference in $F$ between the two data sets below $X \simeq -2$ is significant or due to unknown systematic experimental errors.

\begin{figure}
 \includegraphics[width=8.5cm]{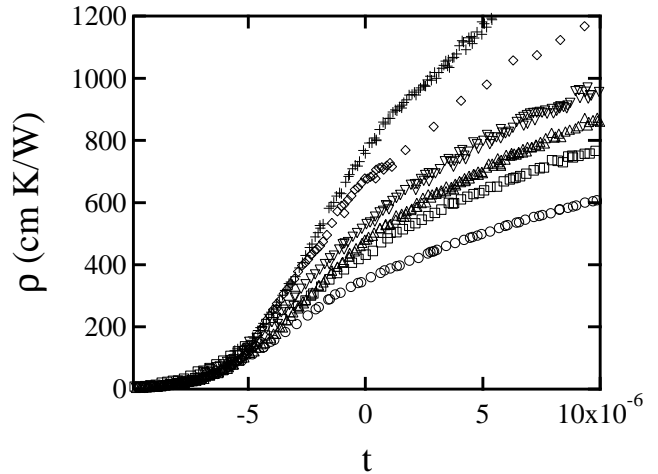}
 \caption{\label{fig:rhoP} Thermal resistivity versus reduced temperature for $L=1.0\mu$m at SVP (open circles); $6.95$bars (open squares);  $11.25$bars (triangles); $14.73$bars (inverted triangles); $22.31$bars (diamonds); $28.00$bars (crosses).  The reduced temperature for each pressure is defined relative to $T_\lambda(P)$.}
 \end{figure}

Figure~\ref{fig:rhoP} shows $\rho(t,P,L)$ as a function of $t$ for six different values of $P$ and $L = 1.0\mu$m. The resistivity does not drop to zero below $t=0$, as is the case for the bulk fluid \cite{tam:a:1985:1}.  The value of $\rho(t=0,P,L)$ varies by nearly a factor of three for the pressures used.

\begin{figure}
 \includegraphics[width=8.5cm]{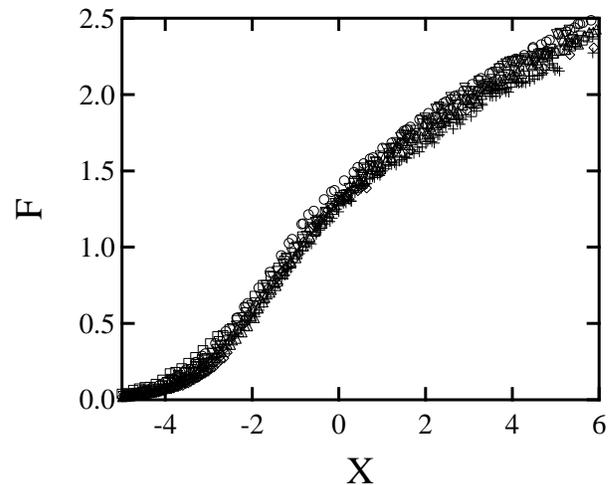}
 \caption{\label{fig:FvsXP} Scaling function $F$ versus scaling variable $X$ for $L=1.0\mu$m at SVP (open circles); $6.95$bars (open squares);  $11.25$bars (triangles); $14.73$bars (inverted triangles); $22.31$bars (diamonds); $28.00$bars (crosses).}
 \end{figure}

In Fig.~\ref{fig:FvsXP} the function $F$ is plotted versus $X$ for six different pressures.  Within our resolution the data collapse on the same curve, suggesting that a single scaling function describes all six pressures.  The collapse occurs despite the large variation of $\rho$ at constant $t$.  The values of $F$ at $X=0$ are given in Table~\ref{tab:F0vsP}.

\begin{figure}
 \includegraphics[width=8.5cm]{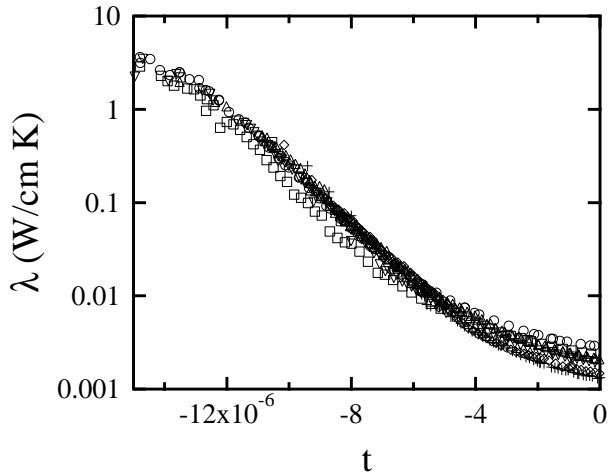}
 \caption{\label{fig:lambdaP} The thermal conductivity below $T_\lambda$ on a logarithmic scale versus reduced temperature on a linear scale for $L=1.0\mu$m at SVP (open circles); $6.95$bars (open squares);  $11.25$bars (triangles); $14.73$bars (inverted triangles); $22.31$bars (diamonds); and $28.00$bars (crosses).   The reduced temperature for each pressure is defined relative to $T_\lambda(P)$.}
 \end{figure}

The thermal conductivity $\lambda$ is plotted on a logarithmic scale versus $t$ on a linear scale in Fig.~\ref{fig:lambdaP} for temperatures  below $T_\lambda(P)$.  It is consistent with an exponential growth  below $T_\lambda(P)$ as noted previously \cite{kahn:a:1995:1}. The amplitudes and arguments of the exponential are approximately the same for all pressures.  These results suggest that $\lambda$, rather than the scaling function $F$, is independent of pressure in this temperature region, and that universality breaks down below $T_\lambda(P)$.

\begin{figure}
 \includegraphics[width=8.5cm]{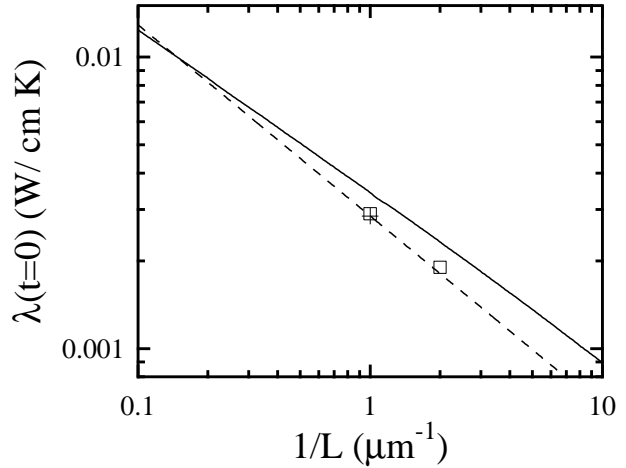}
 \caption{\label{fig:lambdavsLinv} Thermal conductivity $\lambda(t=0)$ vs $L^{-1}$ on logarithmic scales.  The plus is the SVP measurement from Cell I. Open squares are data from cell II. The dashed straight line is the prediction based on Eq.~\ref{eq:F}. The solid curve is the prediction by T\"opler and Dohm \protect\cite{topler:d:2002:1}. }
 \end{figure}

Aside from testing scaling and universality, an important issue is to what extent detailed theoretical calculations can reproduce the conductivity.  As discussed above, the theoretical information is limited. Monte Carlo calculations, which give the shape of the scaling function quite well, involve as yet undetermnined parameters. However, the recent renormalization group calculations have yielded results for $\lambda$ at SVP \cite{topler:d:2002:1}. In Fig.~\ref{fig:lambdavsLinv} we show data for $\lambda(t=0)$ as a function of $L^{-1}$ on logarithmic scales.  The phenomenological scaling function Eq.~\ref{eq:F} predicts $\lambda(t=0) \propto L^{-x/\nu}$ which, for $x/\nu = 0.656$ is shown by the dashed straight line.  The RG prediction is given by the solid line(note that the horizontal axis differs from the original;  here, $L$ is the radius of the channel, not the diameter). Although it comes modestly close to the data points, it does not follow a pure power law, as manifested by curvature of the solid line in the figure. Near $L \simeq 1$ the effective exponent  (i.e. the local slope of the line in Fig.~\ref{fig:lambdavsLinv}) is close to $0.56$, which differs significantly from the prediction based on Eq.~\ref{eq:F}. Although the data points tend to favor the larger exponent $x/\nu$, they do not cover a sufficient range of $L$ to be decisive.  Future plans call for the measurement of the resistivity at larger $L$, with the largest ($50\mu$m) to be flown on the International Space Station, extending the range covered to two decades in $L$.

 \begin{table}
 \caption{\label{tab:F0vsP}Values of the parallel conductance $C_W$ and scaling function $F$ at $X=0$ versus pressure}
 \begin{ruledtabular}
 \begin{tabular}{c|c|c|c|c|c|c|c}
Cell&$L$&$P$&$\xi_0$&$10^{-4}\rho_0$&$x$&$10^4C_W$& $F(0)$\\
&($\mu$m)&(bar)&(nm)&($\textrm{cm}K/W$)&&($W/K$)& \\
\colrule
II & 0.5 & $SVP$ &$0.1432$&$8.312$&$0.4397$& $ 13.3 $ & $1.35$ \\
II & 1 & $SVP$ & $0.1432$&$8.312$&$0.4397$& $ 10.5 $ & $1.35$ \\
I & 1 & SVP & $0.1432$&$8.312$&$0.4397$& $9.35$ & $1.40$ \\
I & 1 & $6.95$ & $0.1425$&$9.073$&$0.4251$& $8.81$ & $1.30$ \\
I & 1 & $11.25$ &$0.1410$&$10.19$&$0.4250$& $8.41$ & $1.30$ \\
I & 1 & $14.73$ &$0.1399$&$11.10$&$0.4250$& $8.07$ & $1.32$ \\
I & 1 & $22.31$ &$0.1382$&$12.79$&$0.4159$& $7.44$ & $1.31$ \\
I & 1 & $28.00$ &$0.1314$&$15.07$&$0.4127$& $6.80$ & $1.24$ \\
 \end{tabular}
 \end{ruledtabular}
 \end{table}

\begin{acknowledgments}
We thank Michael T\"opler and Volker Dohm for sending us their results prior to publication.  This work was supported by NASA Grant No. NAG8-1429.
\end{acknowledgments}



\end{document}